\documentclass[aps,pra,amsmath,showpacs,showkeys,superscriptaddress,twocolumn,10pt]{revtex4-2}
\usepackage{color}
\usepackage{graphicx}
\usepackage{amssymb}
\usepackage{subfigure}
\usepackage{titlesec}
\usepackage{amssymb,amsfonts,amsthm}
\begin{document}
\title{Information theoretic approach to effects of spin-orbit coupling in Bose-Einstein condensates}
\author{Golam Ali Sekh} 
\affiliation{Department of Physics, Kazi Nazrul University, Asansol 713303, India}
\author{Benoy Talukdar}
\affiliation{Department of Physics, Visva-Bharati University, Santiniketan 731235, India}
\author{Supriya Chatterjee}
\affiliation{Department of Physics, Bidhannagar College, EB-2, Sector-1, Salt Lake, Kolkata 700064, India}
\author{Basir Ahamed Khan}
\affiliation{Department of Physics, Krishnath College, Berhampore, Murshidabad 742101, India}
\begin{abstract}
We make use of  Shannon entropy ($S$) and Fisher information ($I$) to study the response of atomic density profiles of a spin-orbit coupled Bose-Einstein condensate to changes in the wave number ($\kappa_L$) of the Raman laser that couples two hyperfine states of atoms in the condensate. The choice for values of $\kappa_L$, the so-called spin-orbit parameter, and Rabi frequency ($\Omega$) leads to two distinct regions in the system's energy spectrum with different order parameters and/or probability densities. In addition, we can have a spatially modulated density profile, reminiscent of the so called stripe phase. Our numbers for $S$ and $I$ demonstrate that for $\kappa_L^2<\Omega$ (region 1) the density profile becomes localized as $\kappa_L$ increases while we observe delocalization in the density distribution for $\kappa_L^2>\Omega$ (region 2) for increasing values of $\kappa_L$. In the stripe phase the nature of $S$ and $I$ to changes in $\kappa_L$ is similar to that found for the condensate in region 2. The results for information theoretic quantities in the stripe phase are, in general, augmented compared to those of region 2. In particular, the highly enhanced values of position-space Fisher information imply an extremely concentrated atomic density distribution to provide an evidence for supersolid properties of Bose-Einstein condensates in the presence of spin-orbit coupling.

\end{abstract}
\maketitle
\section{Introduction}
It is well-known that two popular information measures of a normalized-to-unity probability density $\rho(x)$  are provided by the so-called Shannon entropy \cite{1}
\begin{equation}
 S_{\rho}=-\int_{-\infty}^{\infty}\rho(x)\ln\rho(x)dx
\end{equation}
and Fisher information \cite{2}
\begin{equation}
 I_{\rho}=\int_{-\infty}^{\infty}\rho(x)[\frac{d}{dx}\ln\rho(x)]^2dx.
\end{equation}
In the momentum space results corresponding to the one dimensional quantities in Eqs. (1) and (2) are given by
\begin{equation}
 S_{\gamma}=-\int_{-\infty}^{\infty}\gamma(p)\ln\gamma(p)dp
\end{equation}
and 
\begin{equation}
 I_{\gamma}=\int_{-\infty}^{\infty}\gamma(p)[\frac{d}{dp}\ln\gamma(p)]^2dp,
\end{equation}
where $\gamma(p)$ stands for the normalized-to-unity $p$-space probability density. Bialynicki-Birula and Myceilski \cite{3} introduced a stronger version of the uncertainty relation in terms of position- and momentum-space Shannon entropies. For the one-dimensional system this relation reads
\begin{equation}
S_{\rho}+S_{\gamma}\geq 2.14473. 
\end{equation}
A similar uncertainty relation based on Fisher information is given by \cite{4, 5}
\begin{equation}
I_{\rho}I_{\gamma}\geq4.
\end{equation}
Although both Shannon entropy ($S$) and Fisher information ($I$) are characterized by probability densities corresponding to variation in some observable, $S$ is very little sensitive to changes in the distribution over a small-sized region but $I$ can detect local changes in the distribution. Consequently, these two information measures provide complementary descriptions of disorder in the system. From mathematical viewpoint the former is a convex while the latter is concave \cite{6}. When one grows, the other diminishes. Thus in applicative contexts it will be of interest to make use of the properties of $S$ and $I$ to investigate how does the density distribution of a quantum many-body system respond to external perturbation. In this context there exists a large number of studies \cite{7} on Shannon entropies of many-electron systems which attempt to establish that electron-electron correlation or the so-called Coulomb correlation plays an important role in the delocalization of electrons in the density distribution. More significantly, Romera and Dehesa \cite{8} introduced the product of the position-space Shannon entropy power $(J_\rho=\frac{1}{2\pi e}e^{(2/3)S_{\rho}})$ and Fisher information as an electronic correlation measure of two-electron systems. In the recent past there have  been similar attempts \cite{9, 10} to investigate the  effect of Coulomb correlation in atoms using Fisher information. Rather than electronic cloud in an atom we are interested here in the atomic density profile of a Bose-Einstein condensate (BEC).
\par The experimental realization of BECs \cite{11, 12, 13} led to a progressively growing interest in the physics of cold atoms. The success of the NIST group \cite{14} to generate abelian gauge fields in ultracold atomic systems serves as a typical example. The synthetic magnetic field thus obtained was used to introduce spin-orbit interaction in a BEC consisting of two hyperfine states of $^{87}Rb$ coupled by a Raman Laser. In the present paper we shall make use the information theoretic measures in Eqs. (1) - (4) to examine the role of spin-orbit coupling (SOC) in affecting the density distribution of atoms in a quasi-one dimensional BEC with attractive inter-atomic interaction confined in a harmonic trap.
\par The dynamics of the  SOC BEC is governed by a coupled Gross-Pitaevskii equation (GPE) \cite{15} which, in addition to the parameters of the trapping potential and strength of the inter atomic interaction, relies crucially on the  spin-orbit coupling  parameter ($\kappa_L$) and so-called Rabi frequency ($\Omega$). The parameter $\kappa_L$ is, in fact, the wave number of the Raman Laser that couples two hyperfine atomic states to inject the  effect of spin-orbit coupling in            the condensate. Depending on the choice for the values of $\kappa_L$ and $\Omega$ one can distinguish two different regions in the linear energy spectrum of the system. In region 1, characterized by $\kappa_L^2<\Omega$, the dispersion curve has a single minimum and the associated GPE supports a usual $\sec h^2$ solitonic solution \cite{16}. On the other hand, in region 2 with $\kappa_L^2>\Omega$, the dispersion curve posses two minima at momenta, say, $\pm k_0$, of the system. We can have two different solutions of the GPE corresponding to these minima. In addition, we can have a linear superposition of these solutions that form a stripe phase \cite{17, 18} characterized by a modulated density profile. We shall make use of the density profiles of a quasi-one dimensional SOC BEC corresponding to both regions 1 and 2 to obtain results for  coordinate- and momentum-space Shannon entropy and Fisher information. We shall also compute similar results for condensate in the stripe phase. This will give us an opportunity to examine the role of spin-orbit coupling in localizing or
de-localizing the atomic density distribution in BECs. The spectral properties noted here have also been found to play a role in studying localization properties of the ground state of SOC BEC \cite{19} in optical lattices as well as in controlling Josephson-type oscillation between solitonic components \cite{20} in the BEC.
\par By taking recourse to the use of a multi-scale  expansion method, Achilloes et al \cite{15} provided analytical solutions of the Coupled GPE representing a quasi-one dimensional BEC of our interest. The coordinate-space solution $\overrightarrow{\psi}(x,t)$ giving the upper and lower components of the condensate's order parameter was found to be characterized by a free parameter $\epsilon\sqrt{\omega_0}$. Fortunately, solutions in regions 1 and 2 as well as that representing the stripe soliton are square integrable such that these can be Fourier transformed to get the corresponding momentum-space wave functions $\overrightarrow{\phi}(p,t)$. Consequently, we can construct expressions for normalized-to unity probability densities in both coordinate- and momentum-spaces. In sec. 2 we present results for such probability densities and study their dependence on the tunable spin-orbit coupling parameter \cite{21}. Admittedly, these probability densities in conjunction with Eqs. (1)-(4) provide a useful basis to critically examine how the localization or de-localization in the distribution of cold atoms in a BEC can be controlled by varying the strength of $\kappa_L$. In sec. 3 we present results for Shannon entropies and Fisher information as a function of $\kappa_L$ and try to provide some useful evidence regarding control of atomic density distribution in a SOC BEC by judicious manipulation of the tunable spin-orbit interaction. Finally, we make some concluding remarks in sec. 4.
\section{Density profiles}
Making use of the coordinate-space solution \cite{15} in region 1 and its momentum-space analog we found
\begin{equation}
 \rho_1(x)=\epsilon\sqrt{\frac{\omega_0}{2\Delta}}\sec h^2(\epsilon\sqrt{\frac{2\omega_0}{\Delta}}x)
\end{equation}
and
\begin{equation}
\gamma_1(p)=\frac{\pi}{4\epsilon}\sqrt{\frac{\Delta}{2\omega_0}}\sec h^2(\frac{\pi}{2\epsilon}\sqrt{\frac{\Delta}{2\omega_0}}p)
\end{equation}
for the normalized  position- and momentum-space density distribution for the harmonically trapped cold atoms in our system of interest. Here $\Delta$ is a function of the  spin-orbit coupling constant $\kappa_L$ and Rabi frequency $\Omega$ given by $\Delta=1-\kappa_L^2/\Omega$. The subscript $1$ on $\rho$ and $\gamma$ merely to indicate that these symbols denote the density distribution in region 1. We shall follow the same convention for profiles in region 2. To compute numerical values of $\rho$, $\gamma$ and of $S$, $I$  we shall use $\epsilon^2\omega_0=0.4$ throughout the course of this work. Since the density distribution is real we require $\Delta>0$. In view of this the condition $\kappa_L^2<\Omega$ sets an upper bound for the value of $\kappa_L$.  Here we have chosen to work with $\Omega=150$ and varied from 2 to 10.  In Figs. 1 and 2 we display the plots of $\rho_1(x)$ and $\gamma_1(p)$ as functions of $x$ and $p$ respectively. 
\begin{figure}[h!]
	\begin{center}
		\subfigure[]{\includegraphics[width=8cm, height=6cm]{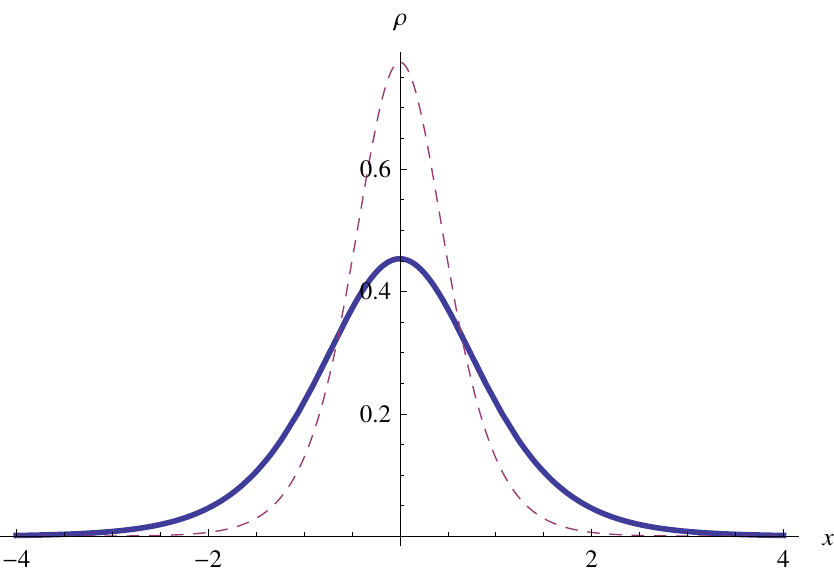}}
		\hspace*{0.5cm}
		\subfigure[]{\includegraphics[width=8cm, height=6cm]{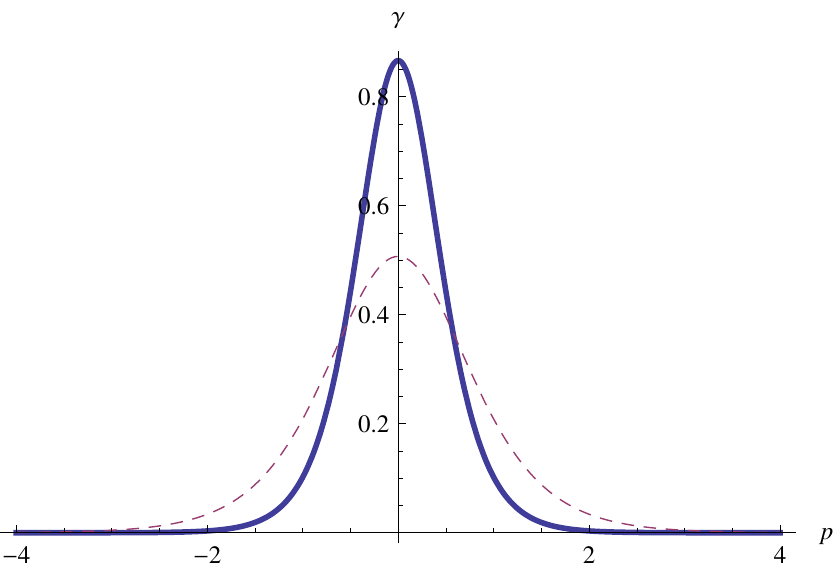}}
		\caption{(a) $\rho_1(x)$ as a function of $x$, (b) $\gamma_1(p)$ as a function of $p$.}
	\end{center}
\end{figure}  
The solid line in Fig.1(a) gives the variation of $\rho_1(x)$ for $\kappa_L=2$ while the dotted curve gives similar variation for $\kappa_L=10$. As compared to solid curve, the dotted curve is highly peaked. This indicates that by increasing the strength of the spin-orbit coupling one can  squeeze the position-space density profile. We have verified that the distribution of cold atoms gradually becomes stiffer as we increase the value of $\kappa_L$. The solid and dotted curves in Fig. 1(b) provide the momentum-space density distribution $\gamma_1(p)$ corresponding to same set of values for $\kappa_L$ as used in Fig. 1(a). As expected, in the reciprocal space, the dotted curve ($\kappa_L=10$) is rather flat compared to the solid curve ($\kappa_L=2$).
\par It is an interesting curiosity to note that, in region 2, the two different position-space order parameters \cite{15} corresponding to the minima $\pm k_0$ of the dispersion curve lead to same normalized density distribution represented by
\begin{equation}
\rho_2(x)=\epsilon\sqrt{\frac{\omega_0}{2}}\frac{\kappa_L}{\kappa_0}\sec h^2(\epsilon\frac{\kappa_L}{\kappa_0}\sqrt{2\omega_0 x}).
\end{equation}
The corresponding momentum-space result reads
\begin{equation}
\gamma_2(p)=\frac{\pi}{4\epsilon\sqrt{2\omega_0}}\frac{\kappa_0}{\kappa_L}\sec h^2(\frac{1}{2\epsilon\sqrt{2\omega_0}}\frac{\kappa_0}{\kappa_L}\pi(\kappa_0+p)).
\end{equation}
Here $\kappa_0=\sqrt{\kappa_L^2-\Omega^2}/\kappa_L$. Since in region 2 $\kappa_L^2>\Omega$, for a chosen value of $\Omega$, we have to use a different set of values of $\kappa_L$ to compute numbers for density profiles. We assume $\Omega=30$ and work with $(\kappa_L)_{min}=6$ and $(\kappa_L)_{max}=14$. In Fig. 2(a) we present the plot of $\rho_2(x)$ as a function of $x$.
\begin{figure}[ht!]
	\begin{center}
		\subfigure[]{\includegraphics[width=8cm, height=6cm]{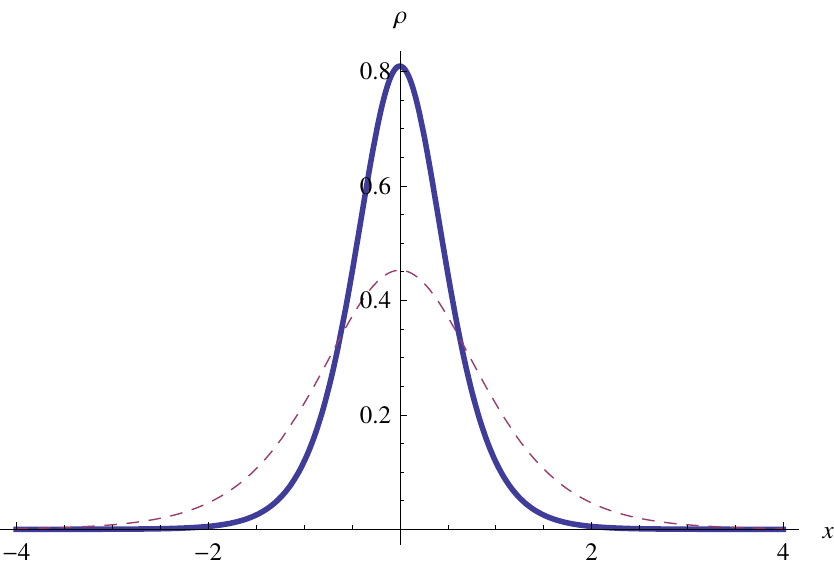}}
		\hspace*{0.5cm}
		\subfigure[]{\includegraphics[width=8cm, height=6cm]{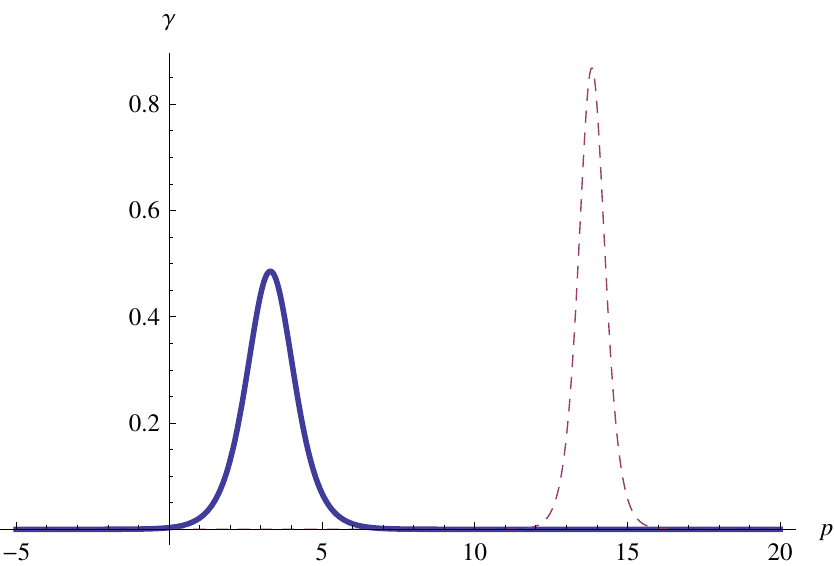}}
		\caption{(a) $\rho_2(x)$ as a function of $x$, (b) $\gamma_2(p)$ as a function of $p$.}
	\end{center}
\end{figure}  
The solid curve gives the variation of $\rho_2(x)$ for $\kappa_L=6$ and the dashed curve portray similar variation for $\kappa_L=14$. Clearly, the density distribution $\rho_2(x)$ is peaked for the lower value of the coupling constant and it becomes flat for higher value of $\kappa_L$. This behavior is just opposite to what we observed for the response of position-space density distribution to spin-orbit coupling. Fig. 2(b) shows the momentum-space density distribution $\gamma_2(p)$ against $p$. Here the distribution represented by the solid curve ($\kappa_L=6$) and that represented by the dashed curve ($\kappa_L=6$) is not overlapping; instead they are situated at two different positions on the $p$ axis. The observed separation, whatsoever, can be attributed to different values of $k_0$ in the argument of the $\sec h^2$ for $\kappa_L=6$ and $14$.
\par In the stripe phase of the condensate the normalized position-space density distribution of cold atoms can be written as
\begin{equation}
 \rho_s(x)=(c_1^2\cos^2(\kappa_0x)+c_2^2\sin^2(\kappa_0x))\sec h^2(bx)/d.
\end{equation}
The momentum-space analog of Eq. (11) is given by
\begin{equation}
 \gamma_s(p)=\frac{\pi}{64bd}(|z_1|^2+|z_2|^2).
\end{equation}
Here, the complex quantities $z_1$ and $z_2$ read
\begin{subequations}
\begin{equation}
z_1=(c_1+c_2)cosec\theta_1\sec\theta_1+(c_1-c_2)(\tan\theta_1+\tan\theta_2^*)
\end{equation}
\mbox{and}
\begin{equation} 
z_2=(c_1-c_2)cosec\theta_1\sec\theta_1+(c_1+c_2)(\tan\theta_1+\tan\theta_2^*)
\end{equation}
\end{subequations}
with
\begin{equation}
 \theta_1=\frac{b+i(\kappa_0+p)\pi}{4b},\;\;\;\theta_2=\frac{b+i(\kappa_0-p)\pi}{4b}.
\end{equation}
The star over $\theta_2$ in Eq. (13) implies complex conjugation. In Eqs. (11) and (12)
\begin{equation}
d=c_1^2+c_2^2+(c_1^2-c_2^2)\kappa_0\pi cosec(\kappa_0\pi/b)/b
\end{equation}
with \cite{15}
\begin{equation}
 c_1=\kappa_L^2+\Omega,\;\;\;c_2=\kappa_0\kappa_L\;\;\;\mbox{and}\;\;\; b=(\epsilon\kappa_L/\kappa_0)\sqrt{2\omega_0}.
\end{equation}
Understandably, the subscript $s$ on $\rho$ and $\gamma$ in Eqs. (11) and (12) has been used to indicate that these quantities refer to density profiles for the condensate in the stripe phase.
\begin{figure}[ht!]
	\begin{center}
		\subfigure[]{\includegraphics[width=8cm, height=6cm]{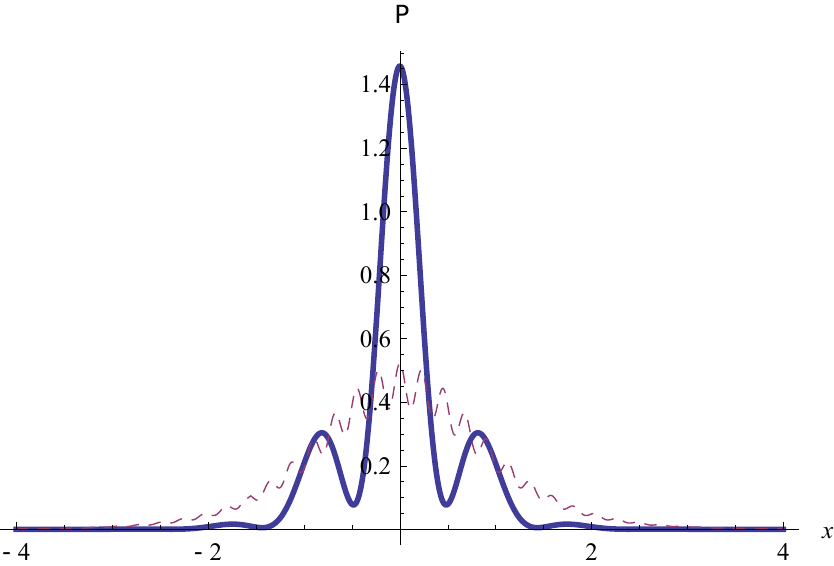}}
		\hspace*{0.5cm}
		\subfigure[]{\includegraphics[width=8cm, height=6cm]{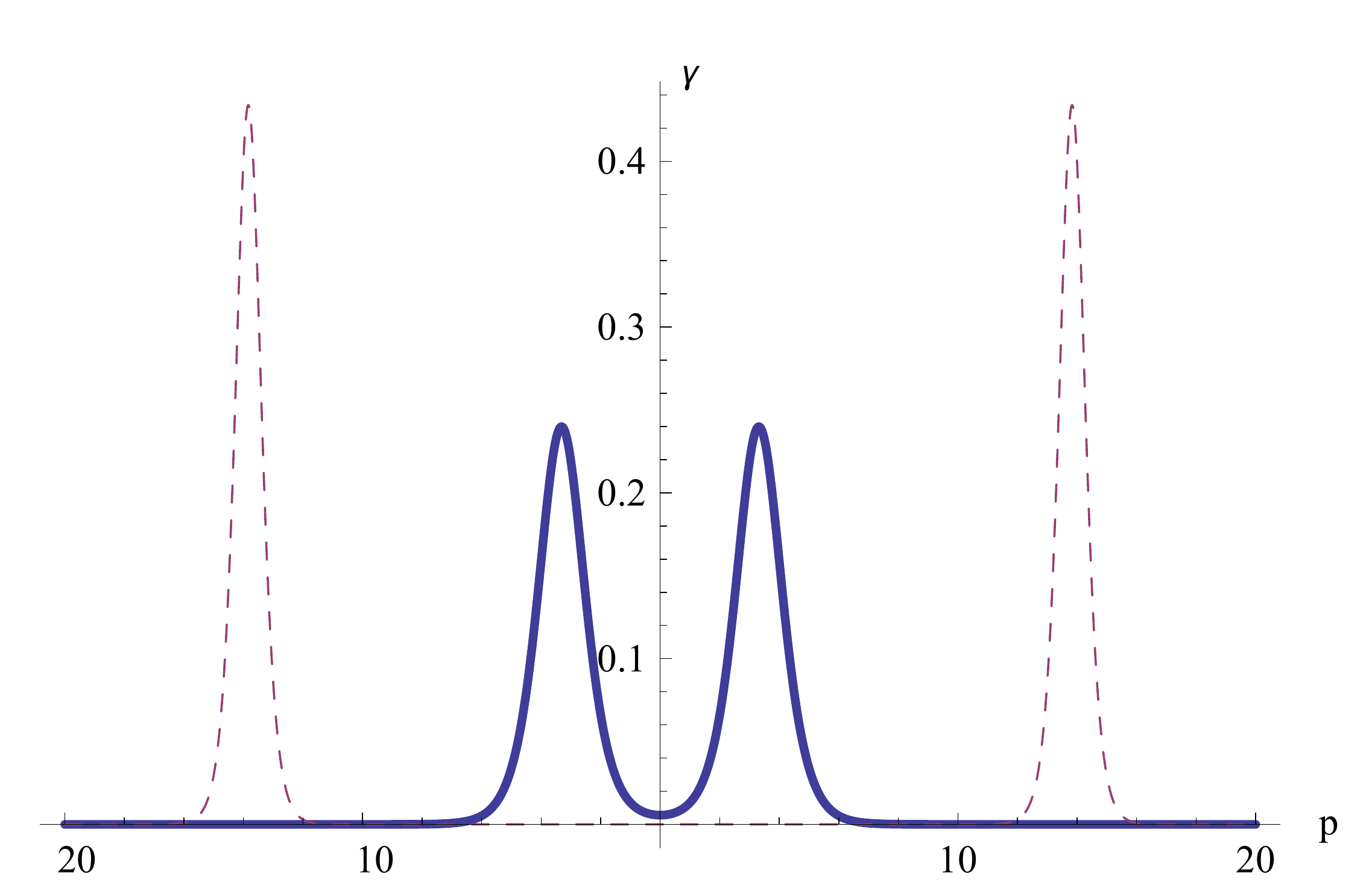}}
		\caption{(a) $\rho_s(x)$ as a function of $x$, (b) $\gamma_s(p)$ as a function of $p$.}
	\end{center}
\end{figure} 
Figures 5 and 6 give the  coordinate- and momentum-density distributions computed by using the expressions in Eqs. (11) and (12). The solid curve in Fig. 3(a) for $\kappa_L=6$, save two humps on the wings, appears to have a $\sec h^2$ shape. We have verified that the number of such humps increases gradually as we go to higher values of the spin-orbit coupling. Finally, as shown by the dashed curve ($\kappa_L=14$) in this figure we have  practically, infinite number of undulations sitting on the curve representing the distribution.  Comparing the curves in Figs. 2(a) and 3(a) we see that the solid curves in both figures are peaked while the dashed curves are relatively flat. The plot of $\gamma_s(p)$ in Fig. 3(b) shows that the solid curve has a saddle-like structure. Clearly, the minimum of this curve at $p=0$ arises due to the maximum of the solid curve in Fig. 3(a) at $x=0$. Similarly, the two maxima in the saddle  can be associated with minima adjacent to the principal maximum of the curve in Fig. 3(a) for $\rho_s(x)$. The dashed curve exhibit two branches placed symmetrically about the $p$ axis and both have $\sec h^2$ shapes. This appears as a point of contrast with the plot of $\gamma_2(p)$ in Fig. 2(b) where there appears only one such dashed curve.
\section{Shannon entropies and Fisher information}
Based on the expression for density profiles  in Eqs. (7) - (12) we shall now present numbers for $S_{\rho}$, $S_{\gamma}$, $I_{\rho}$ and $I_{\gamma}$ for different values of $\kappa_L$ and thus try to gain some physical weight for the effect of spin-orbit interaction on the distribution of cold atoms in SOC BEC.  We present in Table 1 numbers for  information theoretic measures computed by the use of Eqs. (1), (2), (3), (4), (7) and (8). Here we also provide the results for entropy based uncertainties $U_S=S_{\rho}+S_{\gamma}$ and $U_I=I_{\rho}+I_{\gamma}$ respectively. As  used in the plots of  density distributions in Figs. 1 and 2 we work with $\Omega=150$ but vary $\kappa_L$ from 2 to 10 in steps of 2.
\begin{table}[h]
\begin{center}
\begin{tabular}{|l|l|l|l|l|l|l|}
 \hline
$\kappa_L$ & $S_{\rho}$ & $S_{\gamma}$ & $U_{S}$ & $I_{\rho}$ & $I_{\gamma}$ & $U_{I}$\\
\hline
2 & 1.4049 &  0.7572 & 2.1621 & 1.0958 & 4.0027 & 5.0985\\
\hline
4 & 1.3620 & 0.8009 & 2.1629 & 1.1940 &  3.6736 &  4.8676\\
\hline
6 & 1.2812 & 0.8809 & 2.1621 &  1.4035 & 3.1253 &  4.5288\\
\hline
8 & 1.1403 & 1.0218 & 2.1621 & 1.8605 &  2.3577 & 4.2182\\
\hline
10 & 0.8691 & 1.2930 & 2.1621 & 3.2000 & 1.3707 &  4.5707\\
\hline
\end{tabular}
\caption{Shannon entropy and Fisher information for SOC BEC in region 1 ($\kappa_L^2<\Omega$).}
\end{center}
\end{table}
Looking at the numbers for Shannon entropy  we see that $S_{\rho}$ is a decreasing function of $\kappa_L$ and, as expected, in the reciprocal space the results for $S_{\gamma}$ exhibit an opposite behavior.  The Shannon entropy measures the spatial delocalization of a density distribution such that the larger the value of $S_{\rho}$, more delocalized is the density distribution. Thus the observed decrease in the values of $S_{\rho}$ with increasing numbers for $\kappa_L$ implies that by increasing the strength of the spin-orbit constant we go from a diffused to localized atomic distribution in the condensate. From entries in columns  5 and  6 of the table it is seen that the numbers for Fisher information $I_{\rho}$ increase as the spin-orbit coupling becomes stronger and those for $I_{\gamma}$ decrease. This is quite expected since $S$ and $I$ provide complementary descriptions of any probability distribution. More specifically, larger values of $I_{\rho}$ imply more concentrated  density distribution. Thus the results for the Fisher information reconfirm our Shannon-entropy based conclusion regarding  the effect of SOC on atomic density in the BEC.  It is interesting to note the value of $U_{S}$ in column 4 does not depend on $\kappa_L$ and is a constant given by 2.1621. The values of Fisher-information based uncertainty $U_{I}$ in column 7 show some inconsistency in respect of this. However, the values for  and never violate the constraint implied by the uncertainty relations in Eqs. (5) and (6).
\par Table 2 gives the results for Shannon entropies and Fisher information when the parameters of the SOC BEC satisfy conditions of region 2 as identified by the linear energy spectrum of the GPE. All numbers in this table were computed using $\Omega=30$ and varying $\kappa_L$ from 6 to14 in steps of 2. Here $S_{\rho}$ is an increasing function of $\kappa_L$ and $S_{\gamma}$ is a decreasing function. This behavior is just opposite to that exhibited by Shannon entropies in Table 1.  For the lowest value of the spin-orbit coupling parameter, namely, $\kappa_L=6$, the number for is minimum such that the associated distribution is well localized. As we increase the value of $\kappa_L$ the density distribution becomes delocalized. As regards the values of Fisher information we observe that $I_{\rho}$ is a decreasing function of $\kappa_L$ while $I_{\gamma}$ is an increasing function verifying the predictions of Shannon entropies.
\begin{table}[h]
\begin{center}
\begin{tabular}{|l|l|l|l|l|l|l|}
 \hline
$\kappa_L$ & $S_{\rho}$ & $S_{\gamma}$ & $U_{S}$ & $I_{\rho}$ & $I_{\gamma}$ & $U_{I}$\\
\hline
 6 & 0.8256 & 1.3365 & 2.1621 & 3.49091 & 1.25655 & 4.7474\\
\hline
 8 & 1.2944 & 0.8677 & 2.1621 & 1.3670 & 3.2087 & 4.5757\\
\hline
10 & 1.3713 & 0.7908 & 2.1621 & 1.1721 & 3.74223 & 4.9143\\
\hline
12 & 1.3962 & 0.7659 & 2.1621 & 1.1151 & 3.93385 & 5.0489\\
\hline
14 & 1.4066 & 0.7556 & 2.1621 & 1.0923 & 4.01599 &  5.1082\\
\hline
\end{tabular}
\caption{Shannon entropy and Fisher information for SOC  BEC in region 2 ($\kappa_L^2>\Omega$).}
\end{center}
\end{table}
Thus in region 2 by increasing the strength of the spin-orbit coupling we go from a concentrated to delocalized density distribution. This conclusion is supported by values of both Shannon and Fisher information measures. It is interesting to note that the result for $U_S$ in column 4 of Table 2 is exactly the same as the corresponding value in Table 1. Also the values of $U_I$ in both tables are in close agreement with each other. 
\par For the SOC BEC in the stripe phase the results for $S$ and $I$ calculated on the basais of Eqs. (11) and (12)  are displayed in Table 3. Comparing the numbers for $S_{\rho}$ and $S_{\gamma}$ of Table 2 with those in Table 3 we see that as with the results for Shannon entropies for the BEC in region 2, the values of ($S_{\rho}(S_{\gamma})$) for the BEC in the stripe phase also increase(decrease) as $\kappa_L$ increses from 8 to 14. The results for $I_{\rho}$ and $I_{\gamma}$ in both tables take up smaller and larger values with the increasing values of $\kappa_L$.
\begin{table}[h]
\begin{center}
\begin{tabular}{|l|l|l|l|l|l|l|}
 \hline
$\kappa_L$ & $S_{\rho}$ & $S_{\gamma}$ & $U_{S}$ & $I_{\rho}$ & $I_{\gamma}$ & $U_{I}$\\
\hline
6 & 0.27974 & 2.21932 & 2.49906 & 22.52940 & 0.65579 & 23.18519\\
\hline
8 & 1.20534 & 1.58069 & 2.78603 & 27.13030 & 3.08425 & 30.21455\\
\hline
10 & 1.33025 & 1.49086 & 2.82111 & 20.04110 & 3.69123 & 23.73233\\
\hline
12 & 1.37057 & 1.46217 & 2.83274 & 14.79020 & 3.90926 & 18.69946\\
\hline
14 & 1.38736 & 1.45035 & 2.83771 & 11.31680 & 4.00272 & 15.31952\\
\hline
\end{tabular}
\caption{Shannon entropy and Fisher information for tuned SOC-BEC in the stripe phase.}
\end{center}
\end{table}
By comparing the entries in Table 3 for $\kappa_L=6$ with corresponding quantities at higher values of $\kappa_L$, we see that at the lowest value of $\kappa_L$ considered by us the results for $S_{\rho}$ have become rather inconsistent. This is an indication that  stripe phase cannot exist in BECs unless the synthetic spin-orbit coupling is strong enough. It is curious to note that the values of $I_{\rho}$ in Table 3 are greater than the corresponding results in Table 2 by an order of magnitude. But the corresponding numbers for $I_{\gamma}$ in both tables do not differ significantly from each other. Thus the values of $U_{I}$ for the condensate in the stripe phase are much higher than similar numbers for the BEC in region 2. Very high values of $I_{\rho}$ in Table 3 tend to establish that in the stripe phase the density distribution is highly localized. This is quite expected since this phase characterizes the super-solid properties of SOC BEC. On the other hand, from the large values of $U_{I}$ it may be tempting to infer that supersolidity is a purely quantum mechanical phenomenon.
\section{Conclusion}
In this paper we exploited the properties of Shannon entropy and Fisher information to examine the effect of spin-orbit coupling on the density profiles of Bose-Einstein condensates. All results presented  are based on a                     model Hamiltonian for the quasi-one dimensional SOC BEC with attractive inter-atomic interaction confined in a harmonic trap. The energy eigenvalues for the model Hamiltonian of the system come in two different branches defining two separate parametric regions and thus lead to two distinct atomic density profiles. The spinor condensate in region 2 also develops a spontaneous stripe structure to have still a different type of density distribution. The condensate belonging to branch 1 (or region 1) is characterized by $\kappa_L^2<\Omega$ while that corresponding  to branch 2 (or region 2) or stripe phase  is constrained by $\kappa_L^2>\Omega$. The numerical results for $S$ and $I$ indicate that the atomic density distribution in the condensate of region 1 gradually becomes localized as we increase the value of the spin-orbit coupling parameter $\kappa_L$. In contrast to this, the opposite happens for the density profiles of the condensate in region 2 and also for  that in the stripe phase. In both cases the density profiles become delocalized as the value of $\kappa_L$ increases. In general results of $S_{\rho}$, $S_{\gamma}$, $I_{\rho}$ and $I_{\gamma}$ in Table 3 for the condensate in the stripe phase are augmented compared to the corresponding values in Table 2 which displays the information theoretic quantities for the BEC in region 2. More specifically, the values of $I_{\rho}$ in the stripe phase is very large implying that we have here a highly concentrated atomic density distribution. Since BEC is already a superfluid the noted density distribution tends to provide a signature for the condensate to behave like a supersolid \cite{22}.
\par A supersolid represents a state of matter characterized by the coexistence of two spontaneously broken symmetries, namely, the translational symmetry and gauge symmetry \cite{23}. In a recent paper \cite{24} two of us (GAS and BT) have demonstrated that one can try to restore the Galilean variance and/or translational symmetry  of a quasi-one dimensional SOC-BEC by loading the condensate in an optical trap and  periodically modulating the parameters of the nonlinear lattice in the vicinity of the Feshbach resonance. It will, therefore, be quite  interesting to make use of information theoretic methods to study the properties atomic density profiles of an SOC BEC in an optical lattice  with a view to disclose the physics stemming from the interplay between spin orbit coupling and lattice effects. 
\vskip 0.5 cm
\noindent{\bf Acknowledgement}\\
One of the authors (GAS) would like to acknowledge the funding from the 'Science Research and
Engineering Broad, Govt. of India' through Grant No. CRG/2019/000737.

\end{document}